
\centerline {\bf 1. \quad Introduction}
\vskip .20in
It is recognized that primordial
nucleosynthesis provides a unique quantitative window on the early
universe$^{[1,2]}$.
Since the synthesis of
the light elements is determined by events occurring in the epoch from $\sim
1$s to
$\sim 1000$s in
the history of the universe, when the temperatures varied from $\sim 10^{10}
{\rm K}$
$(\ge 1 {\rm MeV})$ to $\sim 10^9 {\rm K} \,\,(\le 0.1 {\rm MeV})$, the
observed abundances constitute a probe of the universe at epochs far earlier
than
those directly probed by the cosmic microwave background radiation (CMBR)
$(t\sim 10^5 $yr;
$T\sim 10^4\,{\rm K} (\sim 1{\rm eV}))$. Thus, through a detailed comparison of
the predicted
abundances with the observational data, proposed cosmological models can be
tested and their controlling parameters can be
constrained. For example, big bang nucleosynthesis (BBN) was found to constrain
the number
of families of light neutrinos$^{[3]}$ prior to accurate accelerator
measurements.

In this paper, we reexplore how the strength of certain primordial magnetic
fields
can be constrained by BBN. If magnetic fields of sufficient
strength existed in the early universe, particularly at or just
before the epoch of primordial nucleosynthesis, they could have had direct
influences on both the
expansion rate of the universe and the nuclear reaction rates.$^{[4,5]}$ These
influences could, of course, affect the
abundances of the light elements produced in this environment. In addition, if
the scale of the primeval
magnetic field were greater than the event horizon, the geometry of the
universe
would also be affected and an anisotropic universe might result. An analysis of
nucleosynthesis in anisotropic Euclidean Universes, in which the dependence of
the primordial abundances of $^4He$, $^3He$, D on the isotropy parameters was
specified more precisely, has been presented by Thorne$^{[4]}$
and by Hawking and Tayler.$^{[6]}$ If a significant degree of anisotropy had
persisted
up to times $\ge 20000$ years, the primordial $^4He$ abundance would have been
reduced to a few per cent. On the contrary,
if the anisotropy is important only during the early stages of the expansion,
the
$^4He$ abundance is about $30\%$ while there is no (negligible) D or $^3He$
production, and one might
hope to eventually reach agreement with the observational values by refining
this
model. This had been done by Juszkiewicz et al, $^{[7]}$ who studied the
influence
of the
anisotropic momentum distribution of neutrinos neglected by Thorne. The
resultant
limit on the
magnetic field at the BBN time, set by the condition of small anisotropy for
$t>1s$ is about $B< 4.1\times 10^{12} {\rm Gauss.}^{[8]}$

On the other hand, if the primeval magnetic field were sufficiently spread over
distances
small compared with the event horizon at that epoch, the geometry of the
universe
would not be affected $^{[9]}$ and it would still be described by a
Robertson-Walker
metric. For this situation, it has been qualitatively pointed out by a number
of
authors$^{[9,10]}$ that, in the presence of a very intense magnetic field
($B\ge 10^{13}$G), the neutron would decay more rapidly than in the field-free
case,
this could obviously affect light element synthesis in a dramatic way.

However, in previous studies, only the effects
of a very strong primordial magnetic field $(B>10^{13}$G) on abundances $^4He$,
D,
and $^3He$, have been studied, not vice-versa, and no critical limit on the
primeval
magnetic field was explicitly derived. The questions we address in this paper
are:
(1) What is the limit on the primordial magnetic field?  (2) How does the
magnetic
field influence the emerging abundances of other light
elements(A$> 9$), such as lithium, boron, etc.?  We find that there are
still constraints to be explored on the strength and coherence scale of
primordial magnetic fields, using observational abundances and BBN.
\vskip .20in
\centerline{\bf 2. \quad The Direct Effects of the Primeval Field on BBN}
\vskip .20in

In the early expansion of the universe, the existence of a large scale
primordial
magnetic
field may have both direct and indirect effects on BBN. In this regard, the two
most
sensitive and competing effects are: (1) the effects on reaction rates and (2)
the
effects on the expansion rate. These two effects will further alter the
resulting abundances of the elements.

For simplicity, we assume that our universe is fully filled by randomly
oriented
and distributed thin-wall magnetic domains (or bubbles)$^{[11]}$. The size of
each
domain is large enough
so that the field inside the domain can be seen as a uniform field, but it
is still small compared with the event horizon. Thus, the magnetic field will
have
similar effects on the motion of particles in each domain. Here, we will
neglect
the boundary effects since the wall is assumed to be thin.

As to the effects on reaction rates, we have recently derived the reaction
rates
as a function of a uniform magnetic field ${\bf \vec B}$ in the presence of an
arbitrary degeneracy and polarization$^{[12]}$. As an application, if we assume
that the magnetic field is nearly uniform in each domain, we can use
our derived results (Ref.[12]) as a first order approximation for our purpose
here.
However, we would like to point out that, if the magnetic field is not uniform
through
the whole universe but varies with spacial variables, or if the scale of the
magnetic
field (or magnetic bubbles) is much smaller than the horizon scale and
the magnetic domains are disconnected from each other, the nuclear reaction
rates will
become inhomogeneous; i.e., the reaction rates will differ from region to
region
even though the geometry of the universe
is still not affected. This would require that we introduce reaction rate
fluctuations
into the standard big bang code (similar to the introduction of the density
fluctuations associated with the first order QCD phase transition$^{[13]}$),
and
perform multizone
calculations. Such exploration is beyond the scope of the present paper and
will be
addressed in future work.

Now let us explore the effects of the magnetic fields on the expansion rate of
the
universe. According to our assumptions, the globally chaotic(but locally
orderly)
magnetic fields will have no effect
on the geometry of the universe. The geometry of the universe is still
described by a
Robertson-Walker metric. For this metric, the work-energy equations can be
expressed
as
$$ {d\over dt}(\rho R^3) + {p\over c^2}{d\over dt}(R^3)=0, \eqno(2.1)$$
where $R(t)$ is the distance measure, $\rho$ is the total mass-energy density,
and $p$
is the total pressure.

In general, we consider that the universe consists of three types of matter
during the epoch of interest. These are: (1) the strongly and
electromagnetically
interacting particles (e.g., nucleons, electrons, photons, etc.), which can be
described
as a perfect fluid; (2) the weakly interacting particles, which nevertheless
affect
the n-p ratio (e.g, electron neutrinos, etc.); and (3) the effectively
noninteracting
particles (e.g., $\nu_\mu$ and so on) which only contribute to the energy
density
but do not enter into specific reactions.
The total mass-energy density $\rho$ and pressure $p$ can be expressed as
$$\rho = \rho_\gamma + \rho_e + \rho_\nu + \rho_b + \rho_B, \qquad p = p_\gamma
+ p_e
+ p_\nu + p_b + p_B, \eqno(2.2)$$
where
$$ \rho_e = \rho_{e^-} + \rho_{e^+}, \qquad p_e = p_{e^-} + p_{e^+},$$
$$\rho_\nu = \rho_{\nu_{e^-}} + \rho_{\nu_{e^+}} + \rho_{\nu_\mu} +
\rho_{\nu_\tau} +
\rho_{\overline\nu_e} + \rho_{\overline{\nu_\mu}} +
\rho_{\overline{\nu_\tau}},$$
$$p_\nu = p_{\nu_{e^-}} + p_{\nu_{e^+}} + p_{\nu_\mu} + p_{\nu_\tau} +
p_{\overline{\nu_e}} + p_{\overline\nu_\mu} + p_{\overline\nu_\tau},$$
and the subscripts $\gamma, \,\, e,\,\, \nu_e,\,\,\nu_\mu,\,\,\nu_\tau,\,\,
b,\,\,
{\rm and}\,\, B$ stand, respectively, for
photons, electrons, e-neutrinos, $\mu$-neutrinos, $\tau$-neutrinos, baryons,
and
magnetic field. Expressions for these
thermodynamic quantities are given below, for the case of nondegenerate
neutrinos$^{[14]}$
$$\eqalign{&\rho_\gamma = 8.42 \, T_9^4 \,{\rm gm\,\, cm^{-3}}, \qquad p_\gamma
= {1\over 3}
\rho_\gamma c^2, \qquad \rho_\nu = 6\,\rho_{\nu_i} = {21\over 8} \rho_\gamma
\big
({T_\nu \over T}\bigr
)^4, \cr
&p_\nu = {1\over 3} \rho_\nu c^2,\quad \rho_e = {7\over 4} \rho_\gamma \,\,(T_9
\gg
6), \quad p_e = {1\over 3} \rho_e c^2\, (T_9 \gg 6),\cr
&\rho_b = 7\times 10^{-6}\, T_9^3\,{\rm gm\, \,cm^{-3}}, \,\, p_b = n_b k T
\sum_i
Y_i,\, Y_i = {X_i \over A_i},\,\, \rho_B = {B^2\over 8\pi}\cr}, \eqno(2.3)$$
where $T_9 = T/(10^9 K),$  $\rho_{\nu_i}\,(\nu_i = \nu_e,\,\,{\overline
\nu_e},\,\,\nu_\mu,\,\,{\overline \nu_\mu},\,\,\nu_\tau,\,\,{\overline
\nu_\tau}\,)\,=
{7\over 16} \rho_\gamma \big({T_\nu \over T}\big )^4$ is the mass density of
each
type of neutrino and antineutrino, $T_\nu$ is the neutrino temperature, $n_b$
is
the number density of baryons, and
$Y_i,$ $A_i$, and $Z_i$ designate the mass fraction, mass number
and atomic number of the ith nucleus.

{}From the assumptions of flux conservation and the presence of a conducting
medium(as
appropriate for the universe prior to recombination),
we can obtain a simple temperature dependence for $B$:
$$\quad B\propto R^{-2} \propto T^2.$$
Therefore, the energy density of the magnetic
field has the same temperature dependence as the energy density of the leptons
and that of the radiation field.

We now define
$$ \rho = \rho_1 + \rho_2,\qquad p = p_1 + p_2,$$
$$\eqalign{&\rho_1 \equiv \rho_\gamma + \rho_{e^-} + \rho_{e^+} + \rho_B,\quad
p_1
\equiv p_\gamma + p_{e^-} + p_{e^+} + p_B,\quad \rho_2 \equiv \rho_\nu +
\rho_b,\quad p_2 = p_\nu + p_b,\cr
&\qquad \qquad \qquad \chi \equiv {\rho_B \over \rho_\gamma + \rho_e + \rho_\nu
+
\rho_b} \equiv {\rho_B \over \rho_0},\qquad \rho_0\equiv \rho(B=0).\cr}
\eqno(2.4)$$
The relation between the magnetic field and radiation is thus
$$\rho_B / \rho_\gamma \simeq {43\over 8} \, \chi. \eqno(2.5)$$
Substituting these into equations (2.2) and (2.1), noticing the following
approximations
$$\rho_b \ll \rho_\nu \to \rho_2 \simeq \rho_\nu,\quad p_b \ll p_\nu \to p_2
\simeq
p_\nu, \eqno(2.6)$$
and using the fact that $\rho_{\nu} \propto R^{-4}$, we obtain
$${dR\over dT} = {-R\over 3\bigr [ \rho_1 (T) + {p_1\over c^2}(T) \bigr ]}
{d\rho_1\over dT}. \eqno(2.7)$$
By using the expansion rate
$${1\over R} {dR\over dt} = \pm \bigr ( {8 \pi G \over 3}\rho \bigr )^{1/2},
\eqno(2.8)$$
where $G$ is the gravitational constant, the relation between the photon
temperature
and the time is found to be
$${dT\over dt} = \mp \bigr ( {8 \pi G \over 3}\rho \bigr )^{1/2} \bigr [ \rho_1
(T) +
{p_1\over c^2}(T) \bigr ]\bigr [ {d\rho_1\over dT}\bigr ]^{-1}. \eqno(2.9)$$
At high temperatures, $T_\nu = T \propto R^{-1}$, and
$$ \rho = \rho_0 ( 1+ \chi ) \simeq {43\over 8} \rho_\gamma (1 + \chi ),
\eqno(2.10)$$
$$ \rho_1 =
{11\over 4} \rho_\gamma + \rho_B = {11\over 4} \rho_\gamma \bigr ( 1 + {43\over
22}
\chi \bigr ) \equiv \rho_{10}\bigr (1 + {43\over 22}\bigr ),\quad \rho_{10}
\equiv
\rho_1(B=0), \eqno(2.11)$$
$${d\rho_1 \over dT} = \bigr ( 1 + {43\over 22} \chi \bigr ) {d\rho_{10} \over
dT} +
{43\over 22} \rho_{10} {d\chi \over dT}, \eqno(2.12)$$
$$ {d\rho_{10} \over dT}\simeq \beta \, T_9^3,\quad \beta = 9.262 \times
10^{-8}.
\eqno(2.13)$$
Incorporating Eq.(2.5) into Eq.(2.8), we obtain a final expression
for the relationship between photon temperature and time
$${dT\over dt} = \mp {\bigr ( {8 \pi G \over 3}\rho_0 \bigr )^{1/2} (1 +
\chi)^{1/2}\rho_{10}\bigr [ 1 + {43\over 22}\chi + {1\over 3} + {p_B \over
\rho_{10}
c^2}\bigr ] \over \bigr (1 + {43\over 22}\chi\bigr )\, \beta T_9^3 + {43\over
22}
\rho_{10} {d\chi\over dT}}. \eqno(2.14)$$
Moreover, considering the fact that, $B \propto T^2, \,\, \rho_B \propto T^4,$
and $
\rho_0 \simeq {43\over 8} \rho_\gamma \propto T^4,$ we can introduce a
convenient invariant measure of magnetic field strength and assume that the
ratio of the magnetic energy density ($\rho_B$) to the total other energy
density ($\rho_0$) is nearly a constant during BBN. This gives ${d\chi\over dT}
= 0$.
We now consider two cases:
\vskip .15in

\noindent
{\it 1. A global zero magnetic pressure} ($p_B = 0$, but $B\not=0$):

Physically, this corresponds to a situation where there exists a non-zero local
uniform
magnetic field (inside each bubble) but a zero total magnetic pressure(pressure
free)
due to the
random distribution of the tangled magnetic bubbles. In this case, Eq.(2.14)
becomes
$${dT\over dt} = \mp {\bigr ( {8 \pi G \over 3}\rho_0 \bigr )^{1/2} (1 +
\chi)^{1/2} \bigr ( 1 + {129\over 88}\chi \bigr ) \over 3 \bigr (1 + {43\over
22}\chi\bigr )}\,T_9. \eqno(2.15)$$
Integrating, we obtain
$$T_9 = \kappa \,{ (1 + {43\over 22}\chi\bigr )^{1/2} \over (1 + \chi )^{1/4}
\bigr ( 1
+ {129\over 88}\chi \bigr )^{1/2}}
t^{-1/2}, \eqno(2.16)$$
where $$\kappa = \big({12 \pi G a\,g_{eff}\over 2 c^2}\big)^{-1/4} \simeq
\cases
{10.4, &if $N_\nu = 2$ and $g_{eff} = 9$;\cr 4.7, &if $N_\nu = 3$ and $g_{eff}
=
{43\over 4}$,\cr}$$
in c.g.s. units, $g_{eff}$ is the "effective" number of relativistic degrees
of freedom(helicity states), and  $a$ is the Stefan black-body constant.
\vskip .15in

\noindent
{\it 2. A non-zero magnetic pressure ($p_B = \rho_B c^2)$:}

If the magnetic field inside each bubble and the distribution of the magnetic
bubbles are not so chaotic, for example, if each magnetic bubble is
dipole-like,
we will have an averaged magnetic pressure $p_B \propto \rho_B c^2$.
In this instance, Eq.(2.14) becomes
$${dT\over dt} = \mp {\bigr ( {8 \pi G \over 3}\rho_0 \bigr )^{1/2} (1 +
\chi)^{1/2} \bigr ( {1\over 3} + {43\over 44}\chi \bigr ) \over 4 \bigr (1 +
{43\over
22}\chi\bigr )}\,T_9 \eqno(2.17)$$
and integration yields
$$T_9 = \kappa \,{ (1 + {43\over
22}\chi\bigr )^{1/2} \over (1 + \chi )^{1/4} \bigr ( 1 + {129\over 44}\chi
\bigr
)^{1/2}} t^{-1/2}, \eqno(2.18)$$

Note that in the limit when the magnetic fields are absent or very weak
($\rho_B =0;\,\, \chi=0,\, {\rm or}\,\, \chi\ll1)$,
Eqs.(2.16) and (2.18) both reduce to
$$T_9 \simeq \kappa \, t^{-1/2}. \eqno(2.19)$$
This is just the formula used in standard BBN
calculations.$^{[14,15]}$

If the magnetic field is very strong, $\chi\gg 1$,
then equation (2.16) and equation (2.18) become, respectively,
$$T_9 = \sqrt{4\over 3} \chi^{-1/4}\,\kappa t^{-1/2},\quad {\rm and} \quad T_9
=
\sqrt{2\over 3} \chi^{-1/4}\,\kappa t^{-1/2}, \eqno(2.20)$$
The dependences of the temperature on the time $t$ and the magnetic parameter
$\chi$,
in the presence of a strong magnetic field, are shown in Figures 1a and
1b. These relations clearly indicate that the effect of the presence
of a strong magnetic field on the expansion rate of the universe are indeed
significant.

Now we introduce another magnetic parameter $\gamma = B/(2B_c)$, where $B_c =
{m_e^2
c^3\over e\hbar} = 4.414\times 10^{13}$gauss is the field strength where
quantized
cyclotron line effects begin to occur; we will refer to this as a quantum
critical
field value$^{[16]}$ (See Appendix). According to our definition of the factor
$\chi$, we have
$$\chi\equiv {B^2 /8\pi \over 43 \rho_\gamma /8} = {B^2 \over 43 \pi a T^4}
\simeq
0.76 \, {\gamma^2 \over T_{10}^4}, \eqno(2.21)$$
where $a$ is the Black-Body constant and $T_{10} = T/(10^{10}\, K)$. If we use
the
``critical" temperature $T_c\sim 1.28 \times 10^{10}K$(note: $nkT_c\sim
B_c^2/8\pi$,
$k$ is Boltzmann constant, $n \simeq 20 T^3$ is number density of particles.).
Eq.(2.21) can be re-expressed as
$$\chi \simeq 0.283\,\bigr ({T_c\over T}\bigr)^4\, \gamma^2. \eqno(2.22)$$
\vskip .20in
\centerline {\bf 3. \quad Numerical Results and a Limit on the Field Strength}
\centerline {\bf and Field Coherence Length}
\vskip .20in

We will now take into account the two independent effects of magnetic fields on
reaction rates and on expansion rates
calculated in Ref. [12] and Section 2, respectively, in a reexamination of big
bang
nucleosynthesis. We use the new version of the Wagoner code developed by
Kawano$^{[16]}$. Specifically,
we have replaced the old formulae in the code, for both the reaction rates and
the expansion rate, with the new derived equations
(3.8), (3.9), (3.10), (3.11), in Ref.[12], and (2.18) above, and
calculated the abundances. These are to be compared with observational data, to
determine the implied constraints on
the strength and coherence scales of a primordial magnetic field. The
observed abundances used are those
summarized by Walker et al 1991$^{[2]}$. The main technique used is to adjust
$\gamma$ until the calculated abundances no longer match the observational
data.

In order to obtain a limit on the strength of primordial magnetic fields
and to focus on the effects on BBN explicitly, we have
fixed all model parameters other than $\gamma$ in our calculations: in
particular, the
neutron lifetime $\tau_n$, the number of neutrino species $N_\nu$, and the
baryon to
photon ratio  $\eta\equiv {n_b\over n_\gamma}$. For our purpose,
we adopt the following values for these parameters$^{[2]}$:
$$\tau_n = 889.6 \pm 2.9 s, \qquad N_\nu =3, $$
$$2.8 \times 10^{-10}\le \eta \le 4.0\times 10^{-10}.\eqno(3.1)$$
Moreover, we have assumed non-degenerate neutrinos ($\phi_e = 0, \phi_\nu = 0)$
(Thomas, Olive, and Schramm$^{[17]}$, Steigman and Kang$^{[18]}$). For these
choices,
we then compute the primordial abundances numerically. Our computational
results are displayed in both tables 1-3 and figures 2-4. Each figure contains
seven
sub-figures ((a), (b), (c), (d), (e), (f), and (g)), which represent the
abundances of
the elements for different strengths of the primordial magnetic fields on a
coherence scale of $L$ ($<$ the event horizon at that epoch $\sim 2\times
10^{12}$cm).

As expected, our calculations reveal that the abundances of the light elements
can be
dramatically affected by a strong magnetic field. For instance, if the magnetic
fields
on scales less than the horizon are as strong as  $B \ge 10^{13}$ gauss, the
abundances
of most elements (except for protons) increase manifestly. In particular, the
concentrations of $^2H$, $^4He$, $^6Li$, $^7Li$,
$^9Be$, $^{14}N$, all are enhanced. Some elements, for example, $^4He$, $^7Li$,
$^8Li$,
$^{11}B$, $^{12}C$, $^{13}C$, $^{14}C$,
and $^{15}N$, show sharp increases (i.e., $^4He \ge 0.5$, ${^7Li / H} \ge 8.43
\times
10^{-9} \sim 10^{-8}$, ${^8Li / H} \ge 7.5 \times 10^{-14} \sim 10^{-13}$,
${^{11}B /
H} \sim 10^{-15}$, etc.), as illustrated
in figures Ia-Id (I=2,3,4) and tables 1-3. On the
contrary, if the magnetic fields are as weak as $10^{11}$ gauss, the
emerging abundances of
the light elements, according to our calculations, are only affected slightly
and the
variations become negligible. Figures Ie-Ig (I=2,3,4) display these
outcomes.

For comparison, we have also computed the effects on the
abundances resulting from the energy density of the magnetic field only
(without any
variations on the reaction rates). We find that the effects on the abundances
of the light elements from reaction rates dominate the contributions from the
energy density, unless the magnetic field is very intense ($B > 10^{13}$
Gauss).
These results are shown in Fig.5 and Fig.6, respectively. Also, as is true for
the
standard BBN model, a high ratio $n_b/n_\gamma$ will globally enhance the high
A
element abundances.

We note that the observed mass fraction of helium is approximately (Skillman
1993)$^{[19]}$
$$Y^{\rm obs}_p = 0.235 \pm 0.01.\eqno(3.2)$$
For other elements, the adopted primordial abundances are displayed in Table 4
(Walker et al 1991)$^{[2]}$. By comparing
the predicted abundances in the presence of B-fields with the observational
determinations, (using particularly the abundance of $^4He$), we can constrain
the
B-field on scales less than the horizon. Figures I(a,b,c,d,e,f,g)
(I = 2,3,4) and tables 1-3 show the abundances of the elements in
the presence of magnetic fields, with field strengths ranging from zero
to $B = 8.8\times 10^{14}$
gauss. Through the comparison between our numerical calculations and the
observational results, we ascertain that to keep the abundances of
light elements compatible with the observations, the primordial magnetic
fields at the BBN epoch ($\sim 1{\rm min.}$) must satisfy the requirement
$$\gamma\leq 0.001, \qquad B \leq 10^{11}G, \eqno(3.3)$$
on scales less than the horizon. At this limit, the calculated abundances of
light
elements are shown in Table 5.

Incorporating our above upper limit into equations (2.21) and (2.5), we can
further estimate the ratio of the energy density between magnetic fields
and cosmic radiations at the BBN epoch $T_{10} \sim 0.1$ as
$${\rho_B\over \rho_\gamma} \sim 4\%. \eqno (3.4)$$

\vskip .15in
We now come to the comparison of the diffusion time of magnetic fields with the
expansion time
of the universe. The characteristic diffusion time for magnetic fields on a
scale
$l(\sim L\,\big({R\over R_{\rm nuc}}\big))$ is $^{[20]}$
$$\tau_d \simeq {4\pi l^2 \sigma \over c^2}.\eqno(4.1)$$
Here $R$ is the cosmological scale factor, and $L$ and $R_{\rm nuc}$ are,
respectively,
the coherence scale of the B-field and the scale factor at the BBN epoch.
Since the expansion time $\tau_{\rm exp}$ of the universe (or Hubble time) goes
as
 $$\tau_{\rm exp} \propto R^2,$$
we also have
$$\tau_d \sim {4\pi l^2 \sigma \over c^2} {\tau_{\rm exp}\over t_{\rm nuc}}
\sigma
.\eqno(4.2)$$
where $t_{\rm nuc}$ is the time of the BBN epoch,
$\sigma = {c^2\over 4\pi\eta} \, {\rm s^{-1}}$ is the electrical conductivity
of
the universe, $\eta \simeq {4\times 10^{12} Z {\rm ln}\Lambda \over T_e^{3/2}}$
esu
describes the Spitzer resistivity (for electron-proton collisions) due to
electron
collisions with
neutral hydrogen$^{[21]}$, $T_e$ denotes the electron temperature in K, $Z$ is
the mean charge of the plasma, and $\Lambda$ is a so called Coulomb integral,
which
has a typical value around $15\pm5$.

The ratio of the diffusion time to the Hubble time can thus be estimated as
$${\tau_d\over \tau_{\rm exp} }\sim 17.5 \bigr({L\over {\rm cm}}\bigr)^2
\tau_{\rm
exp}^{-3/4},\eqno(4.3)$$
where we have used the approximations: $t_{\rm nuc}\sim 1$s and
$T_{\rm nuc}\sim 10^{10}$K. This ratio corresponding to a given physical
scale is a monotonically decreasing function of time. Therefore scales that can
not dissipate at recombination ($\sim 10^{12}-10^{13}$s)
could not have dissipated earlier. Taking the calculation at recombination, we
find that if the coherence scale of the
magnetic fields at the BBN epoch is larger than $10^4 {\rm cm}$ ($\sim 10^{-6}$
of
the horizon scale) at that time, the
field will not be dissipated prior to recombination. The dependence of the
ratio with the expansion time is shown in Figure 8.
\vskip .20in

At the BBN epoch, based on our numerical results in Section 3, the
primordial magnetic fields (or magnetic bubbles) at the BBN epoch are
constrained as $B\le 10^{11} {\rm gauss}$ on scales of $L\ge 10^4 {\rm cm}$
(and
$L\le 10^{12}$cm). To the
extent that the universe is a good conductor, this primeval field will evolve
to the
recombination
era by relation $B \propto R^{-2}$ (and $L\propto R$). The implied
magnetic field at the time of recombination, prior to the structure formation
of the
universe, would thus be
$$B_{\rm rec} \le 0.1\, {\rm gauss}, \eqno(4.4)$$
coherent on scales of $L_{\rm domain} \ge 10^{10} {\rm cm}$ (and $\le
10^{18}$cm).
On scales much larger than the size of the
magnetic domains (or bubbles), the physical mechanisms driving field generation
are
uncorrelated. To put this in current perspective, Hogan$^{[22]}$ has estimated
with certain assumptions that such a field at recombination would
correspond to a intergalactic field limit today of $\le 7\times 10^{-9}{\rm
G}$.
\vskip .20in
\centerline {\bf 5. \quad Conclusions and Future Work}
\vskip .20in

The effects of the magnetic fields on big bang nucleosynthesis and the
cosmological
expansion rate have been investigated generally in this paper
for coherent and chaotic fields on scales smaller than the event horizon.
An upper limit has been provided on the strength of the primordial magnetic
field
on scales smaller than the event horizon.
Our results show that, in the framework of standard big bang nucleosynthesis,
the maximum strength of the primordial magnetic field on scales greater
than $10^4\,{\rm cm}$ but smaller than the
horizon at the BBN epoch ($\sim 10^{12}$cm), can only
be $10^{11}$ gauss, which implies that the magnetic fields
at recombination time would in principle be no stronger than 0.1 gauss.
Moreover, in our
calculations, we find that, of the two major effects of a primordial magnetic
field, those arising from modification of the reaction rates will dominate
those
arising from modification of the
expansion rate (or B-field energy density), unless the magnetic field is very
intense ($B\gg
10^{13} {\rm gauss}.$)

Finally, we here would like to make two comments:

\noindent{\bf a.} {\it Rates fluctuations (or Inhomogeneous model).}

If the magnetic field is not uniform or the size of the magnetic bubbles is
much
smaller than the horizon scale and
the bubbles are disconnected from each other, the nuclear reaction rates will
become
inhomogeneous; i.e., the reaction rates inside a region will differ
from those outside the region, even though the geometry of the universe is
still not
affected. This would require that we introduce reaction rate fluctuations
into the big bang calculation (similar to the introduction of the density
fluctuations associated with the first order QCD phase transition), and perform
multizone calculations.

\noindent{\bf b.} {\it Anisotropy (or effects on geometry).}

If the size of the magnetic bubbles were larger than the horizon scale, the
effects on
the geometry of the universe would need to be examined, and the
Robertson-Walker
metric would need to be replaced by other metrics. In addition, an anisotropy
of
the universe would
result which might have important galactic consequences. (This has been
explored by
Thorne, 1967, but not with the more extensive network used here.)

A subsequent paper will examine these coherence scales.
\vskip .20in
\centerline {\bf Acknowledgements}
We would like to thank A. V. Olinto for very
helpful discussions. This research was supported in part by the NSF grants
AST-93-96039 and AST-92-17969 at the University of Chicago, in part by
NSF grant AST-90-22629, DOE grant
DE FG0291ER40606 and NASA grant NAGW 1321 at the University
of Chicago, and in part by the DOE and by NASA through grant NAGW 2381 at
Fermilab.

\vskip .30in
\vfill
\eject
\centerline {\bf APPENDIX}
\vskip .20in
\centerline {\bf Quantum Mechanical Considerations}
\vskip .20in
The applicability
of classical electrodynamics to electrons requires that the wave-length of the
synchrotron line in the presence of the magnetic
field be much larger than the distances of order of $\hbar/m_e c$, for which
classical electromagnetics will break
down because of quantum effects$^{[23]}$. Let us now derive the limit at
which classical electrodynamics leads to internal contradictions$^{[24]}$.

Consider a system in which a charge $e$, with mass $m$, moves in a uniform
magnetic
field $B$. The synchrotron frequency $\omega$ of motion ( angular frequency )
can
then be written as
$$\omega = e\, c\, B\,/E,\quad {\rm or} \quad \lambda \sim {mc^2\over eB},
\eqno(A.1)$$
where $E$ is the total energy of the charged particle and $\lambda$ is the
wavelength
of the synchrotron line.

If quantum effects become important, we would have
$$E= m c^2 = \hbar \omega, \quad {\rm or }\quad \lambda \sim {\hbar\over
mc}.\eqno(A.2)$$

The magnetic field will then satisfy
$$B \sim B_c = {\omega_c m c^2\over e c} = {m^2 c^3 \over e \hbar} = 4.4 \times
10^{13}
{\rm G}, \eqno(A.3)$$
which presents a limit where quantum electrodynamics plays a role. Therefore,
Eq.(A.3) could be considered in some sense as a quantum critical
limit for an organized primordial magnetic field on large scales. For this
critical
value, we can estimate
the corresponding temperature by letting
$$B_c^2 /8 \pi = n k T_c\,\,, \quad n\simeq 20 T^3 \, K^{-3}, \eqno(A.4)$$
thus
$$ T_c = \bigr ( {B_c^2 \over 160 \pi k}\bigr )^{1/4} \simeq 1.28 \times
10^{10} \,
K.\eqno(A.5)$$
Note that $T_c$ is comparable to the temperature immediately prior to the BBN
epoch.
This means that, if the primordial magnetic field prior to the
BBN epoch were to be as strong as $B>10^{13}$gauss, then it's origin would
probably be
some quantum process in the early universe. For such conditions,
we would not have a field
organized on a large scale as the universe expanded out of the quantum domain,
because
most of the energy released would be converted into small-wave-length radiation
rather
than into an ordered magnetic field. It is interesting that our numerical
calculations ($B<10^{11}$gauss) in Section 3 appear to have ruled this
possibility
out. Instead, it could be suggested
that any primordial magnetic field must have been initially in the classical
regime
$B<10^{11}$gauss.
\vskip .20in
\vfill
\eject
\centerline{\bf REFERENCES}
\vskip .20in
\par \ref
[1] D. N. Schramm and R. V. Wagoner, Ann. Rev. Nucl. \& Part. Sci., (1977).
\par \ref
[2] T. P. Walker, G. Steigman, D. N. Schramm, K. A. Olive, and H-S Kang,
Astrophysics. J., 376 (1991) 51. P.24.
\par \ref
[3] G. Steigman, D. N. Schramm, and J. Gunn, Phys. Lett. B 66 (1977) 202; J.
Yang, M.
S. Turner, G. Steigman, D. N. Schramm, and K. A. Olive, Astrophys. J., 281
(1984) 493.
\par \ref
[4] K. S. Thorne, Astrophys. J. 148 (1967) 51.
\par \ref
[5] Ya. B. Zel'dovich and I. D. Novikov, Relativistic Astrophysics, Vol II The
Structure and Evolution of the Universe, (Chicago: University of Chicago Press,
1983)
\par \ref
[6] S. W. Hawking and R. J. Tayler, Nature, 209 (1966) 1278.
\par \ref
[7] R. Juszkiewicz, S. Bajtlik, and K. Gorski, Mon. Not. R. Astron. Soc. 204
(1983) 63.
\par \ref
[8] Ya. B. Zel'dovich, Astron. J. (SSSR), 60 (1969) 656.
\par \ref
[9] G. Greenstein, Nature, 223 (1969) 938.
\par \ref
[10] J. D. Barrow, Mon. Not. R. Astro. Soc., 175 (1976) 379; R. F. O'Connell
and J. J.
Matese, Nature, 222 (1969) 649.
\par \ref
[11] T. Tajima, S. Cable, K. Shibata, and R. M. Kulsrud, Astrophys.J. 390
(1992) 309.
\par \ref
[12] B. Cheng, D. N. Schramm, and J. W. Truran, Submitted to Phys. Lett. B.
\par \ref
[13] H. Hurki-Suonio, R. A. Matzner, K. A. Olive, and D. N. Schramm, Astrophys.
J.,
353 (1990) 406; J. H. Applegate, C. J. Hogan, and R. J. Scherrer, Phys. Rev. D,
35
(1987) 1151.
\par \ref
[14] R. Wagoner, Astrophys. J. Suppl. series 162 Vol 18 (1969) 247.
\par \ref
[15] P. J. E. Peebles, Astrophys. J., 146 (1966) 542; R. V. Wagoner, W. A.
Fowler, and
F. Hoyle, Astrophys. J., 148 (1966) 3.
\par \ref
[16] L. Kawano, nuc123 big bang code.
\par \ref
[17] Thomas, Olive, and Schramm, Astrophys. J., 406 (1993) 569.
\par \ref
[18] Steigman and Kang, Nucl. Phys. B, 372 (1992) 494.
\par \ref
[19] Skillman, Preprint.
\par \ref
[20] T. G. Cowling, Proc. Roy. Soc. London A, 183 (1945) 453.
\par \ref
[21] L. Spitzer, Jr., Physics of Fully Ionized Gases, (New York: Interscience:
1962); Ya. B. Zel'dovich and I. D. Novikov, Relativistic Astrophysics, Vol 2
(Chicago:
University of Chicago Press, 1983), p.200.
\par \ref
[22] C. J. Hogan, Phys. Rev. Lett., 51 (1983) 1488.
\par \ref
[23] H. Euler and B. Kockel, Naturewissenschaften, 23 (1935) 246.
\par \ref
[24] L. D. Landau and E. M. Lifshitz, The Classical Theory of Fields (Addison -
Wesley
Publishing Company, Inc., Reading Mass., 1965), revised 2nd ed., Chap. 9.
\vfill
\eject
\centerline{\bf Figure Captions}
\vskip .28in

\noindent Fig. 1a The dependence of $T_9$ with t under strong B-field but
$P_B=0$.

\noindent Fig. 1b The dependence of $T_9$ with t under strong B-field but
$P_B\not=0$.

\noindent Fig. 2 Abundance of elements (n - $^8Li$) in the presence of B-fields
but non
degenerate neutrinos.

\noindent Fig. 3 Abundance of the elements ($^8B$ - $^{12}C$) in the presence
of
B-fields but nondegenerate neutrinos.

\noindent Fig. 4 Abundance of the elements ($^{12}N$ - $^{15}O$) in the
presence of
B-fields but nondegenerate neutrinos.

\noindent Fig. 5 Abundance of the elements affected by energy density of
B-field($\gamma = 10$) only.

\noindent Fig. 6 Abundance of the elements affected by energy density of
B-field($\gamma = 100$) only.

\noindent Fig. 7 The ratio of diffusion time of magnetic fields to Hubble time.

\vfill
\eject
\centerline{\bf Table Captions}
\vskip .28in

\noindent Table 1 Abundance of elements (n - $^8Li$) in the presence of
B-fields
but nondegenerate neutrinos.

\noindent Table 2 Abundance of the elements ($^8B$ - $^{12}C$) in the presence
of
B-fields but nondegenerate neutrinos.

\noindent Table 3 Abundance of the elements ($^{12}N$ - $^{15}O$) in the
presence of
B-fields but nondegenerate neutrinos.

\noindent Table 4 Observed abundances.

\noindent Table 5 The calculated abundances at limit $B\le 10^{11}G$.

\vfill
\eject

\end